\documentclass[10pt,letterpaper,english]{article}
\usepackage{times}
\usepackage[T1]{fontenc}
\usepackage[latin1]{inputenc}
\usepackage{babel}
\usepackage{longtable}

\makeatletter

\providecommand{\LyX}{L\kern-.1667em\lower.25em\hbox{Y}\kern-.125emX\@}

 \newenvironment{lyxcode}
   {\begin{list}{}{
     \setlength{\rightmargin}{\leftmargin}
     \raggedright
     \setlength{\itemsep}{0pt}
     \setlength{\parsep}{0pt}
     \fontsize{8}{9.6}\selectfont\ttfamily}%
    \item[]}
   {\end{list}}

\makeatother
\begin{document}

\title{Subclassing errors, OOP, and practically checkable rules to prevent
them}

\author{Oleg Kiselyov\\
Software Engineering, Naval Postgraduate School, Monterey, CA 93943\\
oleg@pobox.com, oleg@acm.org
\date{}
}

\maketitle
\begin{abstract}
This paper considers an example of Object-Oriented Programming (OOP)
leading to subtle errors that break separation of interface and implementations.
A comprehensive principle that guards against such errors is undecidable.
The paper introduces a set of \emph{mechanically verifiable} rules
that prevent these insidious problems. Although the rules seem restrictive,
they are powerful and expressive, as we show on several familiar examples.
The rules contradict both the spirit and the letter of the OOP. The
present examples as well as available theoretical and experimental
results pose a question if OOP is conducive to software development
at all.

\end{abstract}

\section{Introduction}

Decoupling of abstraction from implementation is one of the holy grails
of good design. Object-oriented programming is claimed to be conducive
to such a separation, and therefore to more reliable code. In the
end, productivity and quality are the only true merits a programming
methodology is to be judged upon. This article will discuss a simple
example that questions if Object-Oriented Programming (OOP) indeed
helps separate interface from implementation. First we demonstrate
how easily subclassing errors arise and how difficult (in general,
undecidable) it is to prevent them. We later introduce a set of expressive
rules that preclude the subclassing errors, and can be mechanically
verified. Incidentally the rules run contrary to the OOP precepts.

We take a rather familiar example that illustrates the difference
between subclassing and subtyping: the example of Sets and Bags. The
example is isomorphic to that of circles vs. ellipses or squares vs.
rectangles. Section 2 introduces the example and carries it one step
further, to a rather unsettling result: a \char`\"{}transparent\char`\"{}
change in an implementation suddenly breaks client code that was written
according to public interfaces. We set out to follow good software
engineering practices; this makes the resulting failure even more
ominous. Section 3 brings up a subclassing vs. subtyping dichotomy
and the Liskov principle of behavioral substitutability. We show that
Sets and Bags viewed as mutable or immutable \emph{objects} are not
subtypes of each other. The indiscriminate use of implementation inheritance
indeed prevents separation of interface and implementation. In Section
4 we take a contrary point of view, of bags and sets as values without
a hidden state and whose responses to external messages cannot be
overridden. We prove that a set truly \emph{is-a} bag; a set \emph{is}
substitutable for a bag, a set can always be manipulated as a bag,
a set maintains every invariant of a bag -- and it is still a set.
The section also shows that if we abide by practically checkable rules
we obtain a guarantee that the subtle subclassing errors cannot occur
in principle. We will also show that the rules do not diminish the
power of a language.

Inheritance and encapsulation, two staples of OOP, make checking of
the Liskov Substitution Principle (LSP) for derived objects generally
undecidable. On the other hand, the proposed rules, which \emph{can}
be checked at compile time, make derived values satisfy LSP.

The article aims to give a more-or-less \char`\"{}real\char`\"{} example,
which we can run and see the result for ourselves. By necessity the
example had to be implemented in some language. The present article
uses C++. It appears however that similar code and similar conclusions
can be carried on in many other object-oriented languages (e.g., Java,
Python, etc).

\section{Coupling of interface and implementation}

Suppose I was given a task to implement a Bag -- an unordered collection
of possibly duplicate items (integers in this example). I chose the
following interface:

\begin{lyxcode}
typedef~int~const~{*}~CollIterator;~~~~~//~\textrm{Primitive~but~will~do}

class~CBag~\{

~public:

~~int~size(void)~const;

~~int~count(const~int~elem)~const;

~~virtual~void~put(const~int~elem);

~~virtual~bool~del(const~int~elem);

~~CollIterator~begin(void)~const;

~~CollIterator~end(void)~const;

~

~~CBag(void);

~~virtual~CBag~{*}~clone(void)~const;

~private:~...~~~~~~~~//~implementation~details~elided

\};
\end{lyxcode}
The class \texttt{CBag} defines usual methods to determine the number
of all elements in a bag, to count the number of occurrences of a
specific element, to put a new element into a bag and to remove one.
The latter function returns \texttt{false} if the element to delete
did not exist. We also define the standard enumerator interface \cite{STL}
-- methods \texttt{begin()} and \texttt{end()} -- and a method to
make a copy of the bag. Other operations of the \texttt{CBag} package
are implemented without the knowledge of \texttt{CBag}'s internals:
the print-on \texttt{operator <\,{}<}, the union (merge)
operator \texttt{+=}, and operators to compare CBags and to determine
their structural equivalence. These functions use only the public
interface of the \texttt{CBag} class:

\begin{lyxcode}
void~operator~+=~(CBag\&~to,~const~CBag\&~from);

bool~operator~<=~(const~CBag\&~a,~const~CBag\&~b);

inline~bool~operator~>=~(const~CBag\&~a,~const~CBag\&~b)

\{~return~b~<=~a;~\}

inline~bool~operator~==~(const~CBag\&~a,~const~CBag\&~b)

\{~return~a~<=~b~\&\&~a~>=~b;~\}
\end{lyxcode}
The complete code of the whole example is available in \cite{Code}.
It has to be stressed that the package was designed to minimize the
number of functions that need to know details of CBag's implementation.
Following good practice, I wrote validation code (file \texttt{vCBag.cc}
\cite{Code}) that tests all the functions and methods of the CBag
package and verifies common invariants.

Suppose you are tasked with implementing a Set package. Your boss
defined a set as an unordered collection where each element has a
single occurrence. In fact, your boss even said that a set is a bag
with no duplicates. You have found my CBag package and realized that
it can be used with few additional changes. The definition of a Set
as a Bag, with some constraints, made the decision to reuse the CBag
code even easier.

\begin{lyxcode}
class~CSet~:~public~CBag~\{

~public:

~~bool~memberof(const~int~elem)~const

~~\{~return~count(elem)~>~0;~\}

~

~~//~Overriding~of~CBag::put

~~void~put(const~int~elem)

~~\{~if(!memberof(elem))~CBag::put(elem);~\}

~

~~CSet~{*}~clone(void)~const

~~\{~CSet~{*}~new\_set~=~new~CSet();

~~~~{*}new\_set~+=~{*}this;~return~new\_set;~\}

~~CSet(void)~\{\}

\};
\end{lyxcode}
The definition of a CSet makes it possible to mix CSets and CBags,
as in \texttt{set += bag;} or \texttt{bag += set;} These operations
are well-defined, keeping in mind that a set is a bag that happens
to have the count of all members exactly one. For example, \texttt{set
+= bag;} adds all elements from a bag to a set, unless they are already
present. On the other hand, \texttt{bag += set;} is no different than
merging a bag with any other bag. You too wrote a validation suite
to test all \texttt{CSet} methods (newly defined as well as inherited
from a bag) and to verify common expected properties, e.g., \texttt{a+=a
\( \equiv  \) a}.

In my package, I have defined and implemented a function that, given
three bags \texttt{a}, \texttt{b}, and \texttt{c}, decides if \texttt{a+b}
is a subbag of \texttt{c}:

\begin{lyxcode}
bool~foo(const~CBag\&~a,~const~CBag\&~b,~const~CBag\&~c)

\{~~~~~~~~~~~~~~~~~~~~//~Clone~a~to~avoid~clobbering~it

~~CBag~\&~ab~=~{*}(a.clone());

~~ab~+=~b;~~~~~~~~~~~//~ab~is~now~the~union~of~a~and~b

~~bool~result~=~ab~<=~c;

~~delete~\&ab;

~~return~result;

\}
\end{lyxcode}
It was verified in the test suite. You have tried this function on
sets, and found it satisfactory. 

Later on, I revisited my code and found my implementation of \texttt{foo()}
inefficient. Memory for the \texttt{ab} object is unnecessarily allocated
on heap. I rewrote the function as

\begin{lyxcode}
bool~foo(const~CBag\&~a,~const~CBag\&~b,~const~CBag\&~c)

\{

~~CBag~ab;

~~ab~+=~a;~~~~~~~~~~//~Clone~a~to~avoid~clobbering~it

~~ab~+=~b;~~~~~~~~~~//~ab~is~now~the~union~of~a~and~b

~~bool~result~=~ab~<=~c;

~~return~result;

\}
\end{lyxcode}
It has exactly the same interface as the original \texttt{foo()}.
The code hardly changed. The behavior of the new implementation is
also the same -- as far as I and the package CBag are concerned. Remember,
I have no idea that you are re-using my package. I re-ran the validation
test suite with the new \texttt{foo()}: everything tested fine.

However, when you run your code with the new implementation of \texttt{foo()},
you notice that something \emph{has} changed! The complete source
code \cite{Code} contains tests that make this point obvious: Commands
\texttt{make vCBag1} and \texttt{make vCBag2} run validation tests
with the first and the second implementations of \texttt{foo()}. Both
tests complete successfully, with the identical results. Commands
\texttt{make vCSet1} and \texttt{make vCSet2} test the CSet package.
The tests -- other than those of \texttt{foo()} -- all succeed. Function
\texttt{foo()} however yields markedly different results. It is debatable
which implementation of \texttt{foo()} gives truer results for CSets.
In any case, changing internal algorithms of a pure function \texttt{foo()}
while keeping the same interfaces is not supposed to break your code.
What happened?

What makes this problem more unsettling is that both you and I tried
to do everything by the book. We wrote a safe, typechecked code. We
eschewed casts. g++ (2.95.2) compiler with flags \texttt{-W} and \texttt{-Wall}
issued not a single warning. Normally these flags cause g++ to become
very annoying. You did not try to override methods of CBag to deliberately
break the CBag package. You attempted to preserve CBag's invariants
(weakening a few as needed). Real-life classes usually have far more
obscure algebraic properties. We both wrote validation tests for our
implementations of a CBag and a CSet, and they passed. And yet, despite
all my efforts to separate interface and implementation, I failed.
Should a programming language or the methodology take at least a part
of the blame? \cite{Ousterhout, Does-oo-sync, Cardelli}

\section{Subtyping vs. Subclassing}

The breach of separation between CBag's implementation and interface
is caused by CSet design's violating the Liskov Substitution Principle
(LSP) \cite{LSP}. CSet has been declared a subclass of CBag. Therefore,
C++ compiler's typechecker permits passing a CSet object or a CSet
reference to a function that expects a CBag object or reference. However,
it is well known \cite{Inheritance-not-subtyping} that a CSet is
not a \emph{subtype} of a CBag. The next few paragraphs give a simple
proof of this fact, for the sake of reference. 

The previous section considered bags and sets from the OOP perspective
-- as objects that encapsulate state and behavior. Behavior means
an object can accept a message, send a reply and possibly change its
state. From this point of view, bags and sets are not subtypes of
each other. Indeed, let us define a Bag as an object that accepts
two messages: \texttt{(send~a-Bag~'put~x)} puts an element \texttt{x}
into the Bag, and \texttt{(send~a-Bag~'count~x)} gives the occurrence
count for \texttt{x} in the Bag (without changing \texttt{a-Bag}'s
state). Likewise, a Set is defined as an object that accepts two messages: \texttt{(send~a-Set~'put~x)}
puts an element \texttt{x} into \texttt{a-Set} unless it was already
there, \texttt{(send~a-Set~'count~x)} gives the count of occurrences
of \texttt{x} in \texttt{a-Set} (which is always either 0 or 1). Throughout
this section we use a different, concise notation to emphasize the
general nature of the argument.

Let us consider a function

\begin{lyxcode}
(define~(fnb~bag)~(send~bag~'put~5)~(send~bag~'put~5)~(send~bag~'count~5))
\end{lyxcode}
The behavior of this function, its contract, can be summed as follows:
given a Bag, the function adds two elements into it and returns \texttt{(+
2 (send orig-bag 'count 5))}. Technically you can pass to \texttt{fnb}
a Set object as well. Just as a Bag, a Set object accepts messages
\texttt{'put} and \texttt{'count}. However applying \texttt{fnb} to
a Set object will break the function's post-condition stated above.
Therefore, passing a set object where a bag was expected changes the
behavior of a program. According to the LSP, a Set is not substitutable
for a Bag -- a Set cannot be a subtype of a Bag. 

Let us consider another function

\begin{lyxcode}
(define~(fns~set)~(send~set~'put~5)~(send~set~'count~5))
\end{lyxcode}
The behavior of this function is: given a Set, the function adds an
element into it and returns 1. If you pass to this function a bag
(which -- just as a set -- replies to messages \texttt{'put} and \texttt{'count}),
the function \texttt{fns} may return a number greater than 1. This
will break \texttt{fns}'s contract, which promised always to return
1.

One may claim that \char`\"{}A Set is not a Bag, but an ImmutableSet
is an ImmutableBag.\char`\"{} This is not correct either. An immutability
per se does not confer subtyping to \char`\"{}derived\char`\"{} classes
of data, as a variation of the previous argument shows \cite{web-site}.
C++ objects are record-based. Subclassing is a way of extending records,
with possibly altering some slots in the parent record. Those slots
must be designated as modifiable by a keyword \texttt{virtual}. In
this context, prohibiting mutation and overriding makes subclassing
imply subtyping. This is the reasoning behind BRules introduced below.
However merely declaring the state of an object immutable is not enough
to guarantee that derivation leads to subtyping: An object can override
parent's behavior without altering the parent. This is easy to do
when an object is implemented as a functional closure, when a handler
for an incoming message is located with the help of some kind of reflexive
facilities, or in prototype-based OO systems \cite{web-site}. Incidently,
if we do permit a derived object to alter its base object, we implicitly
allow behavior overriding. For example, an object \texttt{A} can react
to a message \texttt{M} by forwarding the message to an object \texttt{B}
stored in \texttt{A}'s slot. If an object \texttt{C} derived from
\texttt{A} alters that slot it hence overrides \texttt{A}'s behavior
with respect to \texttt{M}.

The OOP point of view thus leads to a conclusion that neither a Bag
nor a Set are subtypes of the other. The interface or an implementation
of a Bag and a Set appear to invite subclassing of a Set from a Bag,
or vice versa. Doing so however will violate the LSP -- and we have
to brace for strikingly subtle errors. The previous section intentionally
broke the LSP to demonstrate how insidious the errors are and how
difficult it may be to find them. Sets and Bags are very simple types,
far simpler than the ones that typically appear in a production code.
Since LSP when considered from an OOP point of view is undecidable,
we cannot count on a compiler for help in pointing out an error. As
Section 2 showed, we cannot rely on validation tests either. We have
to \emph{see} the problem \cite{Does-oo-sync, Ousterhout, Cardelli}.

\section{Mechanically preventing subclassing errors}

Bags and sets -- as objects -- indeed are not subtypes. Subclassing
them violates LSP, which leads to insidious errors. Bags and sets
however do not have to be viewed as objects. We can take them as pure
values, without any state or intrinsic behavior -- just like the numbers
are. In Section 2, CBag and CSet objects encapsulated a hidden state
-- a collection of integers. The objects had an ability to react to
messages, e.g., \texttt{put} and \texttt{del}, by altering their state.
In this section we re-do the example of Section 2 using a different
approach. Bags and sets no longer have a state that is distinct from
their identity and that can be altered. Equally important we do not
allow any changes to the behavior of bags and sets with respect to
applicable operations, by overriding or otherwise. In other words,
every post-condition of a bag or a set constructor holds throughout
the lifespan of the constructed values. This approach makes the subclassing
problems and breach of encapsulation disappear. It turns out that
a set truly \emph{is-a} bag; a set is substitutable for a bag, a set
can always be manipulated as a bag, a set maintains every invariant
of a bag -- and it is still a set. 

The LSP says, {}``If for each object o1 of type S there is another
object o2 of type T such that for all programs P defined in terms
of T, the behavior of P is unchanged when o1 is substituted for o2,
then S is a subtype of T.{}'' If type T denotes a set of values that
carry their own behavior, and if values of type S can override some
of T values behavior, the LSP is undecidable. Indeed, a mechanical
application of LSP must at least be able to verify that all methods
overridden in S terminate whenever the corresponding methods in T
terminate. This is generally impossible. On the other hand, if T denotes
a set of (structured) data values, and S is a subset of these values
-- e.g., restricted by range, parity, etc. -- the LSP is trivially
satisfied.

This section also shows that if one abides by mechanically verifiable
rules he obtains a guarantee that the subtle subclassing errors cannot
occur in principle. The rules do not reduce the power of a language.

\subsection{BRules}

Suppose I was given a task to implement a Bag -- an unordered collection
of possibly duplicate items (integers in this example). This time
my boss laid out the rules, which we will refer to as \emph{BRules}:

\begin{itemize}
\vspace{-5pt}
\item no virtual methods or virtual inheritance
\vspace{-5pt}
\item no visible members or methods in any public data structure (that is,
in any class declared in an \texttt{.h} file)
\vspace{-5pt}
\item no mutations to public data structures

\begin{itemize}
\vspace{-5pt}
\item a strict form: no assignments or mutations whatsoever
\vspace{-5pt}
\item a less strict form: no function may alter, directly or indirectly,
any data it receives as arguments
\end{itemize}
\end{itemize}
The rules break the major tenets of OOP: for example, values no longer
have a state that is separate from their identity. Prohibitions on
virtual methods and on modifications of public objects are severe.
It appears that not much of C++ is left. Surprisingly I still can
implement my assignment without losing expressiveness -- and perhaps
even gaining some. The exercise will also illustrate that C++ does
indeed have a pure functional subset \cite{Stroustrup}, and that
you can program in C++ without assignments.

\subsection{Interface and implementation of a FBag}

\begin{lyxcode}
class~FBag~\{

~public:

~~FBag(void);

~~FBag(const~FBag\&~another);~~~~//~Copy-constructor

~~\textasciitilde{}FBag(void);

~

~private:

~~class~Cell;~~~~~~~~~~~~~~~~~~~//~Opaque~type

~~const~Cell~{*}~const~head;

~~FBag(const~Cell~{*}~const~cell);~//~Private~constructor

~~~~~~~~~~~~~~~~~~//~Declaration~of~three~friends~elided

\};
\end{lyxcode}
Indeed, there are no virtual functions, no methods or public members.
We also declare functions that take a FBag as (one of the) arguments
and return the count of all elements or a specific element in the
bag, print the bag, \( fold \) \cite{fold} over the bag, compare
two bags for structural equivalence, verify bag's invariants, merge
two bags, add or delete an element. The latter three functions do
not modify their arguments; they return a new FBag as their result.
It must be stressed that the functions that operate on a FBag are
not FBag's methods; in particular, they are not a part of the class
FBag, they are not inherited and they cannot be overridden. The implementation
is also written in a functional style. FBag's elements are held in
a linked list of cells, which are allocated from a pre-defined pool.
The pool implements a mark-and-sweep garbage collection, in C++.

Forgoing assignments does not reduce expressiveness as the following
snippet from the FBag code shows; the snippet implements the union
of two FBags:

\begin{lyxcode}
struct~union\_f~\{

~~FBag~operator()~(const~int~elem,~const~FBag~seed)~const~\{

~~~~return~put(seed,elem);

~~\}

\};

FBag~operator~+~(const~FBag\&~bag1,~const~FBag\&~bag2)

\{

~~return~fold(bag1,union\_f(),bag2);

\}
\end{lyxcode}
Following good practice, I wrote a validation code (file \texttt{vFBag.cc}
\cite{Code}) that tests all the functions of the FBag package and
verifies common invariants.

\subsection{Implementation of a FSet. FSet is a subtype of a FBag}

Suppose you are tasked with implementing a Set package. Your boss
defined a set as an unordered collection where each element has a
single occurrence. In fact, your boss even said that a set is a bag
with no duplicates. You have found my FBag package and realized that
it can be used with few additional changes. The definition of a Set
as a Bag (with some constraints) made the decision to reuse the FBag
code even easier.

\begin{lyxcode}
class~FSet~:~public~FBag~\{

~public:

~~FSet(void)~\{\}

~~FSet(const~FBag\&~bag)~:~FBag(remove\_duplicates(bag))~\{\}

\};

~

bool~memberof(const~FSet\&~set,~const~int~elem)

\{~return~count(set,elem)~>~0;~\}
\end{lyxcode}
Surprisingly, this is the \emph{whole} implementation of a FSet. A
set is fully a bag. Because FSet constructors eventually call FBag
constructors and do no alter the latter's result, every post-condition
of a FSet constructor implies a post-condition of a FBag constructor.
Since FBag and FSet values are immutable, the post-conditions that
hold at their birth remain true through their lifespan. Because all
FSet values are created by an FBag constructor, all FBag operations
automatically apply to an FSet value. This concludes the proof that
an FSet is a \emph{subtype} of a FBag.

The \texttt{FBag.cc} package \cite{Code} has a function \texttt{verify(const
FBag\&)} that checks to make sure its argument is indeed a bag. The
function tests FBag's invariants, for example:

\begin{lyxcode}
const~FBag~bagnew~=~put(put(bag,5),5);

assert(~count(bagnew,5)~==~2~+~count(bag,5)~\&\&

~~~~~~~~size(bagnew)~==~2~+~size(bag)~);

assert(~count(del(bagnew,5),5)~==~1~+~count(bag,5)~);
\end{lyxcode}
Your validation code passes a non-empty set to this function to verify
the set is indeed a bag. You can run the validation code \texttt{vFSet.cc}
\cite{Code} to see for yourself that the test passes. On the other
hand, FSets do behave like Sets:

\begin{lyxcode}
const~FSet~a112~=~put(put(put(FSet(),1),1),2);

assert(~count(a112,1)~==~1~);

~

const~FSet~donce~=~FSet()~+~a112;

const~FSet~dtwice~=~donce~+~a112;

assert(~dtwice~==~a112~);
\end{lyxcode}
where \texttt{a112} is a non-empty set. The validation code \texttt{vFSet.cc}
you wrote contains many more tests like the above. The code shows
that a FSet is able to pass all of FBag's tests as well as its own.
The implementation of FSets makes it possible to take a union of a
set and a bag; the result is always a bag, which can be made a set
if desired. There are corresponding test cases as well. 

To clarify how an FSet may be an FBag at the same time, let us consider
one example in more detail:

\begin{lyxcode}
~~~~~//~An~illustration~that~an~FSet~is~an~FBag

int~cntb(const~FBag~v)~\{

~~FBag~b1~=~put(v,~5);~~FBag~b2~=~put(b1,~5);

~~FBag~b3~=~del(b2,~5);

~~return~count(b3,~5);~\}

const~int~cb1~=~cntb(FBag());~//~cb1~has~the~value~of~1

const~int~cb2~=~cntb(FSet());~//~cb2~has~the~value~of~1

~

~~~~~//~An~illustration~that~an~FSet~does~act~as~a~set

int~cnts(const~FSet~v)~\{

~~FSet~s1~=~put(v,~5);~FSet~s2~=~put(s1,~5);

~~FSet~s3~=~del(s2,~5);

~~return~count(s3,~5);~\}

const~int~cs~=~cnts(FSet());~//~cs~has~the~value~of~0
\end{lyxcode}
This example is one of the test cases in \texttt{vFSet.cc} \cite{Code}.
You can run it and check the results for yourself. Yet it is puzzling:
how come \texttt{cs} has the value different from that of \texttt{cb1}
if there is no custom \texttt{del()} function for FSets? The statement~\texttt{FSet
s2 = put(s1, 5);} is the most illuminating. On the right-hand side
is an expression: putting an element 5 to a FBag/FSet that already
has this element in it. The result of that expression is a FBag \{5,5\},
with two instances of element 5. The statement then constructs a FSet
\texttt{s2} from that bag. A FSet constructor is invoked. The constructor
takes the bag \{5,5\}, removes the duplicate element 5 from it, and
\char`\"{}blesses\char`\"{} the resulting FBag to be a FSet as well.
Thus \texttt{s2} will be a FBag and a FSet, with one instance of element
5. In fact, \texttt{s1} and \texttt{s2} are identical. A FSet constructor
guarantees that a FBag it constructs contains no duplicates. As objects
are immutable, this invariant holds forever.

\subsection{Discussion}

Surprising as it may be, assertions \char`\"{}a Set is a Bag with
no duplicates\char`\"{} and \char`\"{}a Set always acts as a Bag\char`\"{}
do not contradict each other, as the following two examples illustrate:\begin{longtable}{|c|c|}
\hline 
\multicolumn{1}{|p{3.1in}|}{Let \texttt{\{value ...\}} be an unordered collection of values: a
Bag. Let us consider the following values: 

\( vA:42,\, vB:\{42\},\, vC:\{43\},\, vD:\{42\, 43\},\, vE:\{42\, 43\, 42\} \)

\( vA \) is not a collection; \( vB \), \( vC \), \( vD \), and
\( vE \) are bags. \( vB \), \( vC \), and \( vD \) are also Sets:
unordered collections without duplicates. \( vE \) is not a Set.
Every Set is a Bag but not every Bag is a Set.}&
\multicolumn{1}{|p{3.1in}|}{Let \emph{uf-integer} denote a natural number whose prime factors
are unique. Let us consider the following values:

\( vA:\frac{5}{4},\, vB:42,\, vC:43,\, vD:1806,\, vE:75852 \)

\( vA \) is not an integer; \( vB \), \( vC \), \( vD \), and
\( vE \) are integers. \( vB \), \( vC \), and \( vD \) are also
uf-integers. \( vE \) is not a uf-integer as it is a product \( 2*2*3*3*7*7*43 \)
with factors 2, 3, and 7 occurring several times. Every uf-integer
is an integer but not every integer is a uf-integer.}\\
\hline 
\multicolumn{1}{|p{3.1in}|}{We introduce operations \( merge \) (infix \texttt{+}) and \( subtract \)
(infix \texttt{-}). Both operations take two Bags and return a Bag.
Either of the operands, or both, may also be a Set. The result, a
Bag, may or may not be a Set. For example,

\begin{description}
\item [\( vB+vC\Rightarrow vD \)]Both of the operands and the result are
also Sets 
\item [\( vB+vD\Rightarrow vE \)]The argument Bags are also Sets, but the
resulting Bag is not a Set 
\item [\( vE+vE\Rightarrow \{42\, 43\, 42\, 42\, 43\, 42\} \)]None of the
Bags here are Sets
\item [\( vD-vC\Rightarrow vB \)]The argument Bags are also Sets, so is
the result.
\item [\( vE-vC\Rightarrow \{42\, 42\} \)]One of the arguments is a Set,
the resulting Bag is not a Set.
\item [\( vE-vE\Rightarrow \{\} \)]The argument Bags are not Sets, but
the resulting Bag is. \end{description}
}&
\multicolumn{1}{|p{3.1in}|}{We introduce operations \( multiply \) (infix {*}) and \( reduce \)
(infix \%): \( a\%b=a/gcd(a,b) \). Both operations take two integers
and return an integer. Either of the operands, or both, may also be
a uf-integer. The result, an integer, may or may not be a uf-integer.
For example,

\begin{description}
\item [\( vB*vC\Rightarrow vD \)]Both of the operands and the result are
also uf-integers
\item [\( vB*vD\Rightarrow vE \)]The argument integers are also uf-integers,
but the resulting integer is not a uf-integer
\item [\( vE*vE\Rightarrow 5753525904 \)]None of the integers here are
uf-integers
\item [\( vD\%vC\Rightarrow vB \)]The argument integers are also uf-integers,
so is the result
\item [\( vE\%vC\Rightarrow 1764 \)]One of the arguments is a uf-integer,
the resulting integer is not a uf-integer 
\item [\( vE\%vE\Rightarrow 1 \)]The argument integers are not uf-integers,
but the resulting integer is.\end{description}
}\\
\hline 
\multicolumn{1}{|p{3.1in}|}{Bags are closed under operation \( merge \) but subsets of Bags --
Sets -- are not \emph{not} closed under \( merge \). On the other
hand, both Bags and Sets are closed under \( subtract \).

We may wish for a merge-like operation that, being applied to Sets,
always yields a Set. We can introduce a new operation: \( merge-if-not-there \).
We can define it specifically for Sets. Alternatively, the operation
can be defined on Bags; it would apply to Sets by the virtue of inclusion
polymorphism as every Set is a Bag. Sets \emph{are} closed with respect
to \( merge-if-not-there \).

On the other hand, to achieve closure of Sets under \( merge \) we
can project -- coerce -- the result of merging of two Sets back into
Sets, a subset of Bags. The FBag/FSet package took this approach.
If we \( merge \) two FSets and want to get an FSet in result we
have to specifically say so, by applying a projection (coercion) operator:
\texttt{FSet}::\texttt{FSet(const FBag\& bag)}. That operator creates
a new FBag without duplicates. This fact makes the latter a FSet.
Thus \( FSet(vB+vD)\Rightarrow vD \), an FSet.}&
\multicolumn{1}{|p{3.1in}|}{Integers are closed under operation \( multiply \) but subsets of
integers -- uf-integers -- are \emph{not} closed under \( multiply \).
On the other hand, both integers and uf-integers are closed under
\( reduce \).

We may wish for a multiply-like operation that, being applied to uf-integers,
always yields a uf-integer. We can introduce a new operation: \( lcm \),
the least common multiple. This operation is well-defined on integers;
it would apply to uf-integers by the virtue of inclusion polymorphism
as every uf-integer is an integer. uf-integers \emph{are} closed with
respect to the \( lcm \) operation.

On the other hand, to achieve closure of uf-integers under \( multiply \)
we can project -- coerce -- the product of two uf-integers back into
uf-integers, a subset of integers. If we \( multiply \) two uf-integers
and want to get a uf-integer in result we have to specifically say
so, by applying a projection (coercion) operator: \( remove-duplicate-factors \).
That operator creates a new integer without duplicate factors. This
fact makes the resulting integer a uf-integer. Thus \( uf-integer(vB*vD)\Rightarrow vD \),
a uf-integer}\\
\hline
\end{longtable}

It has to be stressed that the two columns of the above table are
not merely similar: they are isomorphic. Indeed, the right column
is derived from the left column by the following substitution of words
that preserves meaning:  Bag \( \leftrightarrow  \) integer, Set
\( \leftrightarrow  \) uf-integer,  merge \( \leftrightarrow  \)
multiply, subtract \( \leftrightarrow  \) reduce. The right column
sounds more \char`\"{}natural\char`\"{} -- so should the left column
as integers and uf-integers are representations for resp. FBags and
FSets.

From an extensional point of view \cite{On-understanding-types},
a type denotes a set of values. By definition of a FSet, it is a particular
kind of FBag. Therefore, a set of all FSets is a subset of all FBags:
FSet is a subtype of FBag. A FBag or a FSet do not have any \char`\"{}embedded\char`\"{}
behavior -- just as integers do not have an embedded behavior. Behavior
of numbers is defined by operations, mapping from numbers to numbers.
Any function that claims to accept every member of a set of values
identified by a type T will also accept any value in a subset of T.
Frequently a value can participate in several sets of operations:
a value can have several types at the same time. For example, a collection
\{ 42 \} is both a Bag and a Set. This fact should not be surprising.
In C++, a value typically denoted by a numeral 0 can be considered
to have a character type, an integer type, a float type, a complex
number type, or a pointer type, for any declared or yet to be declared
pointer type. This lack of behavior is what puts FBag and FSet apart
from CBag and CSet discussed in the previous article. FSet is indeed
a subtype of FBag, while CSet is not a subtype of a CBag as CSet has
a different behavior. Incidentally LSP is trivially satisfied for
values that do not carry their own behavior. FBags and FSets are close
to so-called predicate classes. Since instances of FSets are immutable,
the predicate needs to be checked only at a value construction time.

\subsection{Polymorphic programming with BRules}

The FSet/FBag example above showed BRules in the context of subtypes
formed by a restriction on a base type. As it turns out, BRules work
equally well with existential (abstract) types. To illustrate this
point, the source code accompanying this article \cite{Code} contains
three implementations of a collection of polymorphic values. The collection
is populated by Rectangles and Ellipses, which are instances of concrete
classes implementing a common abstract base class Shape. A Shape is
an existential type that knows how to draw, move and resize itself.
A file Shapes-oop.cc gives the conventional, OOP-like implementation,
with virtual functions and such. A file Shapes-no-oop.cc is another
implementation, also in C++. The latter follows BRules, has no assignments
or virtual functions. Any particular Shape value is created by a Shape
constructor and is not altered after that. Shapes-no-oop.cc achieves
polymorphic programming with the full separation of interface and
implementation: If an implementation of a concrete Shape is changed,
the code that constructs and uses Shapes does not even have to be
recompiled! The file defines two concrete instances of the Shape:
a Square and a Rectangle. The absence of mutations and virtual functions
guarantees that any post-condition of a Square or a Rectangle constructor
implies the post-condition of a Shape. Both particular shapes can
be passed to a function     \texttt{set\_dim(const Shape\& shape,
const float width, const float height);} Depending on the new dimensions,
a square can \emph{become} a rectangle or a rectangle square. You
can compile Shapes-no-oop.cc and run it to see that fact for yourself.

It is instructive to compare Shapes-no-oop.cc with Shapes-h.hs, which
implements the same problem in a purely functional, strongly-typed
language Haskell. All three code files in the Shapes directory solve
the same problem the same way. Two C++ code files -- Shapes-oop.cc
and Shapes-no-oop.cc -- look rather different. On the other hand,
the purely functional Shapes-no-oop.cc and the Haskell code Shapes-h.hs
are uncanny similar -- in some places, frighteningly similar. This
exercise shows that BRules do not constrain the power of a language
even when abstract data types are involved.

\section{Conclusions }

It is known, albeit not so well, that following the OOP letter and
practice may lead to insidious errors \cite{Ousterhout, Cardelli}.
Section 2 of this article showed how subtle the errors can be even
in simple cases. In theory, there are rules -- LSP -- that could prevent
the errors. Alas, the rules are in general undecidable and not \emph{practically
reinforceable}.

In contrast, BRules introduced in this article can be statically checked
at compile time. The rules outlaw certain syntactic constructions
(for example, assignments in some contexts, and non-private methods)
and keywords (e.g., \texttt{virtual}). It is quite straightforward
to write a lint-like application that scans source code files and
reports if they conform to the rules. When BRules are in effect, subtle
subclassing errors like the ones shown in Section 2 become impossible.
To be more precise, with BRules, \emph{subclassing implies subtyping}.
Subclassing by definition is a way of creating (derived) values by
extending, restricting, or otherwise specializing other, parent values.
A derived value constructor must invoke a parent value constructor
to produce the parent value. The former constructor often has a chance
to alter the parent constructor's result before it is cast or incorporated
into the derived value. If this chance is taken away, the post-condition
of a derived value constructor implies the post-condition of the parent
value. Disallowing any further mutations guarantees the behavioral
substitutability of derived values for parent values at all times.

As the examples in this article showed, following BRules does not
diminish the power of the language. We can still benefit from polymorphism,
we can still develop practically relevant code. Yet BRules blur the
distinction between the identity and the state, a characteristic of
objects. BRules are at odds with the practice if not the very mentality
of OOP. This begs the question: Is OOP indeed conducive to software
development?

One can argue that OOP -- as every powerful technique -- requires
extreme care: knives are sharp. Likewise, goto is expressive, and
assembler- or microcode-level programming are very efficient. All
of them can lead to bugs that are very difficult, statically impossible,
to find. On the other hand, if you program, for example, in Scheme,
you never have to deal with an \char`\"{}invalid opcode\char`\"{}
exception. That error becomes simply impossible. Furthermore, {}``while
opinions concerning the benefits of OOSD {[}Object-Oriented Software
Development{]} abound in OO literature, there is little empirical
proof of its superiority{}'' \cite{OOP-ups-downs}.

\subsubsection*{Acknowledgments }

I am grateful to Valdis Berzins for valuable discussions and suggestions
on improving the presentation. This work has been supported in part
by the National Research Council Research Associateship Program, Naval
Postgraduate School, and the Army Research Office under contracts
38690-MA and 40473-MA-SP.

\end{document}